\documentclass[preprint2]{aastex}

\slugcomment{}

\shorttitle{Probing Dust around Brown Dwarfs}
\shortauthors{Apai, D. et al.}

\begin{document}


\title{Probing Dust around Brown Dwarfs:\\
 The naked LP 944-20 and the Disk of Cha H$\alpha$2
 \altaffilmark{1}}


\author{D. Apai\altaffilmark{2}, I. Pascucci\altaffilmark{2}, and Th. Henning\altaffilmark{2}}
\affil{Astrophys. Inst. and Univ. Observatory, Schillerg\"asschen 2-3, D-07745 Jena, Germany}
\email{apai@mpia-hd.mpg.de}
\and 
\author{M. F. Sterzik}
\affil{ESO, Alonso de Cordova 3107, Vitacura, Casilla 19001,  Santiago 19, Chile}
\and
\author{R. Klein and D. Semenov}
\affil{Astrophys. Inst. and Univ. Observatory, Schillerg\"asschen 2-3, D-07745 Jena, Germany}
\and
\author{E. G\"unther and B. Stecklum}
\affil{TLS Tautenburg, Sternwarte 5, D-07778 Tautenburg, Germany}

\altaffiltext{1}{Based on observations obtained at the 3.6m/ESO
 Telescope at La Silla (Chile) for the program 68.C-0446 and
 268.C-5750} 
\altaffiltext{2}{Max Planck Institute for Astronomy, K\"onigstuhl 17,  D-69117, Heidelberg, Germany }


\begin{abstract}
We present the first mid-infrared (MIR) detection of a 
field brown dwarf (BD) and the first ground-based MIR measurements 
of a disk around a young BD candidate. We prove the absence of warm dust surrounding the 
field BD LP 944-20.  In the case of the young BD candidate Cha H$\alpha$2, we find 
clear evidence for thermal dust emission from a disk. Surprisingly, the object does 
not exhibit any silicate feature as previously predicted.  We show that the flat 
spectrum can be explained by an optically thick flat dust disk.

\end{abstract}




\section{Introduction}

Brown Dwarfs (BDs) occupy the substellar mass domain. 
Having masses lower than 75 M$_{\rm Jup}$,  they are unable to burn hydrogen steadily. 
Although their presence has been already predicted  in the sixties by \cite{Kumar}, 
their low luminosity delayed their discovery until 1995, when \cite{Nakajima}  announced 
the first detection of a BD orbiting the nearby M-dwarf star Gl229A.
Recently, the large-scale near-infrared (NIR) surveys 2MASS and DENIS -- complemented
by optical data -- substantially increased the number of known field BDs. 
Additionally, deep NIR surveys of star-forming regions  revealed hundreds of  young BD 
candidates. 

In spite of the rapidly growing number of known BDs \citep{Basri}, 
we do not know if they 
form like planets or like stars. Proposed scenarios  include the straightforward 
star-like formation via fragmentation and disk accretion  \citep{Elmegreen}, 
the ejection of stellar embryos \citep{Reipurth} from multiple systems and the
formation in circumstellar disks like giant planets. 
  The  presence of disks and their properties are crucial in distinguishing 
  between the various scenarios: a truncated disk (size of a few AU)
  would support the ejected stellar embryo hypothesis, a 
 non-truncated one is the sign of stellar-like accretion, while BDs formed 
 like planets should have no dust around them. 

In the case of BDs, NIR data are not necessarily a good tracer of disk 
emission because they are strongly affected by molecular bands of the cool BD atmosphere.
Since the emission of warm (100 - 400 K) circumstellar dust peaks 
around 10 \micron ,  mid-infrared (MIR) excess emission -- arising from dust grains close to the star -- 
is the best tool to search for  circumstellar disks. 
The MIR regime is best accessed by space-born telescopes, the last of which 
was the Infrared Space Observatory  (ISO), operating between 1995 - 1998. 
However, the majority of BDs has been discovered too late to be targeted by ISO.

\begin{figure}
\plotone{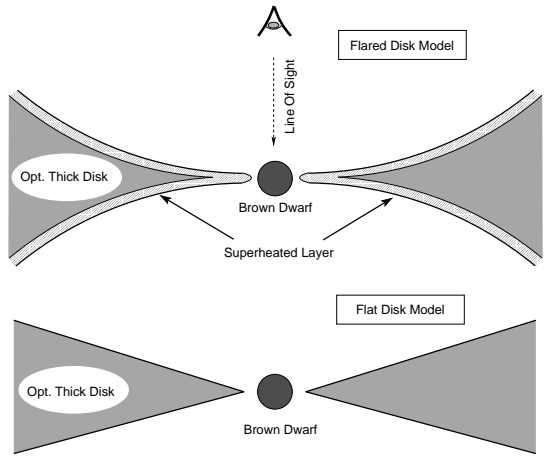}
\caption{Cross sections of the flared and the flat disk model. 
The shaded area represents the optically thin superheated layer 
in the flared disk.  This region is the source of the silicate 
emission feature. The flat disk lacks this disk atmosphere.
\label{Figure1}}
\end{figure}

 Up to now, only few BDs with MIR excess are known. These objects,
identified in the ISOCAM archive, 
 are located in the Cha I or in the $\rho$ Ophiuchi star-forming 
regions \citep{Persi,Comeron98}.
 Their substellar nature has been deduced from comparing NIR and optical 
 measurements to evolutionary models \citep{Comeron98,Comeron}. \cite{NattaTesti}
 proposed a model based on scaled-down disks around pre-main-sequence stars  
 to explain the measured spectral energy distributions (SEDs).
 They followed the Chiang \& Goldreich disk geometry which has been rather 
 successfull in
 describing SEDs of T Tauri and Herbig Ae/Be stars \citep{CG97,Natta, CG01}.
 The main assumption here is the flaring of the disk, which introduces a 
 superheated surface layer, called the disk atmosphere (see Fig.~\ref{Figure1}).
 This optically thin layer produces a strong silicate emission 
feature around 9.7 \micron \ (Si--O stretching mode).  
In an optically thick flat disk, no such feature is expected. 
Therefore, the presence or lack of such a feature is a strong constraint
to any disk model. 

In this paper we present results from our TIMMI2 MIR imaging 
  campaign. Our aim was to detect MIR excess emission and thus to probe
  the presence of warm circumstellar dust around BDs. We targeted 
  8 very close field  BDs of various ages and a young BD candidate in the Cha I 
  star-forming region. Our observations are the first data in the wavelength 
  region of the silicate feature.

\section{Observations}

We carried out deep MIR observations with the 3.6m/ESO Telescope 
at La Silla (Chile) using the TIMMI2 camera \citep{Reimann} 
in 2001 November and December.
The targets were 8 field BDs and a young BD in the Cha I star-forming region.
From the closest field BDs, we selected those which seemed to be the youngest based on
their brightness  and spectral type.

 We used the 9.8 \micron \ filter, where the instrument is the most sensitive, to
search for disk emission. In the case of the detected field BD we also
complemented the 9.8 \micron \ measurement with 5 and 11.9 \micron \ observations;  
the BD detected in the Cha I region was also observed at 11.9 \micron. 
 We applied long exposure times (typically 2 hrs in each filter) in order 
to reach the $\sim$10 mJy sensitivity limit of the instrument. 
Extensive testing of the pointing accuracy shows a typical error 
not larger than 1.5\arcsec{} towards the Cha I star-forming region. This
excludes any confusion with other sources.

\section{Results}
\subsection{Field Brown Dwarfs} 

 Among the 8 targeted nearby field BDs, only the object LP 944-20 could be detected. 
Table \ref{Tab1} summarizes the names and coordinates of the non-detected BDs.
 The 3 $\sigma$ upper limit of the flux density for these sources is 15 mJy
at 9.8 \micron.  
 
 As one of the youngest (475-650 Myr) and closest (5 pc) field BDs 
\citep{Tinney}, LP 944-20 
was the most promising of our targets.
 Based on its optical 
spectrum, its spectral type is equal to or later than M9V \citep{Kirkpatrick}.
Its classification as a BD has been confirmed by the presence of lithium in its photosphere 
\citep{Tinney}. 

  Excellent atmospheric conditions 
and a long integration time led to the {\em first detection 
of a field BD in the MIR}. The fluxes measured  at 5, 9.8 and 11.9 \micron \
are 39 mJy, 24 mJy, and 22 mJy, respectively. These measurements correspond 
to more than 5 $\sigma$ detections in each filter. 
We estimate a photometric error smaller than 15\% for each measurement.

\subsection{Cha H$\alpha$2}

 In contrast to the older 
 field BDs, we found clear evidence for excess MIR emission  
 in the case of the much younger (2 - 4.5 Myr) BD candidate Cha H$\alpha$2.
  The observed fluxes are 17$\pm$2 
 and 21$\pm$3 mJy at 9.8 and 11.9 \micron , respectively.

The object is close to or in the substellar 
domain, depending on its exact age \citep{Comeron}.
 There is some evidence that Cha H$\alpha$2  
is actually a close binary with the components in the substellar domain 
\citep{Neuhauser}.

\section{Discussion}
\subsection{Field Brown Dwarfs}
 The non-detection of the 7 field BDs proves the lack of
 significant amount of warm dust around older field BDs.
These data clearly show that the disk dissipation time is below a few 
100 Myr, consistent with recent measurements of BDs in the young 
$\sigma$ Orionis cluster \citep{Oliveira}.   
 
 Even the detection of the closest target, the 475-650 Myr old LP 944-20, confirms this
 hypothesis. Compared to a simple blackbody with T$_\star$=2300 K,  
 R$_\star$=0.1 R$_{\odot}$, D=5 pc \citep{Tinney}, it is clear that our measurements show no 
 MIR excess, but the photospheric flux of the BD itself.

\begin{figure}
\plotone{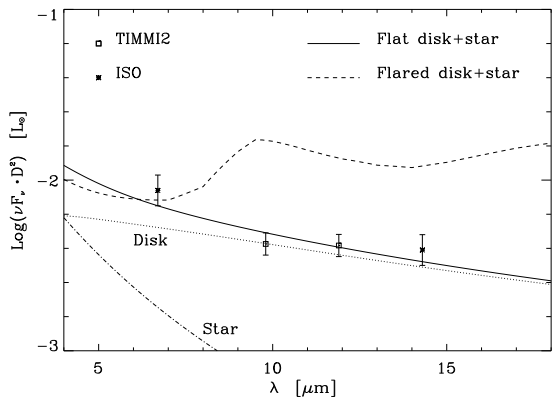}
\caption{ Modelled spectral energy distribution of a flat and a flared disk 
compared to the observations. The asterisks show the ISO measurements at 6.7 
and 14.3 \micron{} with 20\% errorbars, while the squares are 
our TIMMI2 measurements at 9.8 and 
11.9 \micron{}. The estimated errors amount to 15\%. The dot-dashed line 
indicates the contribution from the star, the dotted line is the flat disk 
emission. Their sum is the solid line fitting the observations. The dashed 
line shows the prediction of  the flared disk model.\label{Figure2}}
\end{figure}

\subsection{Disk Model for Cha~H$\alpha$2}
A black body with a temperature and radius corresponding to the classification
of Cha H$\alpha$2 fails to reproduce the observed fluxes that are an 
order of magnitude higher.
 This excess, previously observed by ISO \citep{Persi},
  was interpreted as emerging from a circumstellar disk \citep{Comeron}. 
Our ground-based data corroborate the presence of a disk, 
and adds important constraints to the models. 

The very low 
extinction towards Cha H$\alpha$2 (A$_{\rm I}<$0.4 mag, \citet{Comeron}) 
argues for a nearly face-on disk, where a strong silicate emission feature
would be expected from the superheated layer of a flared disk.
Using the Chiang--Goldreich model \cite{NattaTesti} predict a 10 \micron 
\ emission stronger than 40 mJy. In contrast, we measured fluxes of 17$\pm$2 and 21$\pm$3 mJy 
at 9.8 and 11.9 \micron , respectively. The same flux densities at 
the peak and on the wing of the feature exclude the presence of any silicate 
feature.

In order to understand the reason of the model's deviation from the 
observations,
we inspect its parameters and structure. The Chiang--Goldreich model consists
of three major components: the star, the optically thin disk atmosphere 
and the optically thick disk interior (see Figure \ref{Figure1}).
In the MIR the stellar radiation is well approximated by a black body,
while the optically thick disk emission is given by a power law  
$F_\nu \sim \lambda^{5/3}$ \citep{CG97}.  
 A simple analytical formula is used to describe the optically thin 
disk atmosphere, which is producing the silicate emission \citep{Natta}.

 Changing the disk geometry (inner and outer radii,
scale height, inclination)  is insufficient to explain the observed SED.
 Altering the optical properties (or composition)
of the dust grains has a stronger effect: the absence of the 
 silicate feature could 
 be explained by the lack of this dust component or the presence of 
 large grains (radii larger than 5 \micron).
  However, we stress that the power law continuum 
 $F_\nu \sim \lambda^{5/3}$,  predicted by {\em the flared model,
 does not fit our data}.

A much simpler and more straightforward solution is the assumption 
that the BD is surrounded by an optically thick flat disk. We assume
 a power law disk with a surface density of $\Sigma \propto R^{-3/2}$ and
 a temperature of $T \propto R^{-3/4}$  which are typical of reprocessing 
 and viscous disks \citep{Shu}. Since this disk is entirely optically thick, its
 SED is independent of the dust properties. {\em The model does 
 not show any feature.  The continuum of a power-law flat disk
 has the observed slope}. In Fig. \ref{Figure2} we compare the measurements 
 with model predictions.

\section{Summary}
Our ground-based measurements represent a new  way of probing the properties of 
disks around BDs, exploring their spectral energy distribution and 
therefore constraining model prescriptions. We prove the absence of the 
previously predicted silicate emission feature in the case of the face-on disk 
around the young BD-candidate Cha H$\alpha$2, one of the three known BD-candidates 
with MIR excess. An optically thick flat disk provides a perfect match to our 
data.  Because no evidence for disks around older field BDs could be detected, 
disk dissipation times must be shorter than a few 100 Myr. Our results suggest 
that newborn BDs have disks like young, low-mass stars, but also 
indicate unexpected differences in their disk geometry. The next step in the 
investigation of disks around young BDs is the search for millimetre emission 
to detect the outer parts of the disks and to go to fainter objects with the 
space-based SIRTF mission. 

\acknowledgments
We thank P. Apai, J. Steinacker, 
K. Schreyer and S. Wolf, for their help and advice.

\clearpage


\clearpage

\clearpage
\begin{deluxetable}{lcc}
\tabletypesize{\scriptsize}
\tablecaption{Field Brown Dwarfs undetected in our TIMMI2 campaign.
The 3 $\sigma$ upper flux limit at 9.8 \micron \ is 15 mJy.\label{Tab1}}
\tablewidth{0pt}
\tablehead{
\colhead{ID} &
\colhead{R.~A. (2000)} & 
\colhead{Decl. (2000)}
}

\startdata
2MASSW J2224488-015852  	       & 22 24 43.8 & -01 58 52 \\
2MASSW J00 51 107-15 44 17     & 00 51 10.7 & -15 44 17  \\
2MASSW J03 37 036-17 58 07     & 03 37 03.6 & -17 58 07 \\
DENIS-J-0021-4244	       & 00 21 05.7 &  -42 44 50 \\
DENIS-P-J0255-4700.		       & 02 55 03.3 & -47 00 49.0 \\
2MASSW J0832046-012833         & 08 32 04.6 & -15 48 00.0 \\
\enddata
\tablecomments{Units of right ascension are hours, minutes, 
and seconds, and units of declination are degree, 
arcminutes, and arcseconds.}
\end{deluxetable}



\begin{thebibliography}{}
\bibitem[Basri(2000)]{Basri} Basri, G. 2000, ARA\&A, 38, 485 
\bibitem[Chiang \& Goldreich(1997)]{CG97} Chiang, E. I., \& Goldreich, P. 1997, \apj, 490, 368 
\bibitem[Chiang et al.(2001)]{CG01} Chiang, E. I., Joung, M. K., Creech-Eakman, M. J., 
Qi, C., Kessler, J. E., Blake, G. A., \& van Dishoeck, E. F. 2001, \apj, 547, 1077
\bibitem[Comer\'on et al.(1998)]{Comeron98} Comer\'on, F., Rieke, G. H., Claes, P., Torra, J., \&
Laureijs, R. J. 1998, \aap, 335, 522
\bibitem[Comer\'on et al.(2000)]{Comeron} Comer\'on, F.,  Neuh\"auser, R., \&  Kaas, A. A. 2000, \aap, 359, 269
\bibitem[Elmegreen(1999)]{Elmegreen} Elmegreen, B. G. 1999, ApJ, 522, 915
\bibitem[Kirkpatrick et al.(1997)]{Kirkpatrick} Kirkpatrick, J. D., Henry, T. J., \& Irwin, M. J. 1997, \aj, 113, 1421
\bibitem[Kumar(1963)]{Kumar} Kumar, S. S. 1963,  \apj, 137, 1121 
\bibitem[Nakajima et al.(1995)]{Nakajima}Nakajima, T., Oppenheimer, B. R.,
 Kulkarni, S. R., Golimowski, D. A., Matthews, K., \& Durrance, S. T.  1995, Nature, 378,  463
\bibitem[Natta et al.(2000)]{Natta} Natta, A., Meyer, M. R., \& Beckwith, S. V. W. 2000, \apj, 534, 838
\bibitem[Natta \& Testi(2001)]{NattaTesti} Natta A. \&  Testi, L. 2001, \aap, 376, L22 
\bibitem[Neuh\"auser et al.(2002)]{Neuhauser} Neuh\"auser, R., Brandner, W., Alves, J., Joergens, V., \&  Comer\'on, F. 2002, \aap, 384, 999
\bibitem[Oliveira et al.(2001)]{Oliveira} Oliveira, J. M., Jeffries, R. D., Kenyon, M. J., Thompson, S. A., \& Naylor, T. 2001, \aap, 382, L22
\bibitem[Persi et al.(2000)]{Persi} Persi, P. et al. 2000, \apj, 357, 219-224 
\bibitem[Reipurth \& Clarke(2001)]{Reipurth} Reipurth, G. \&  Clarke, C. 2001, \aj, 122,  432
\bibitem[Reimann et al.(2000)]{Reimann} Reimann, H.-G., Linz, H., Wagner, R., Relke, H.,
 Kaeufl, H. U., Dietzsch, E., Sperl, M., \& Hron, J. 2000, in Optical and IR Telescope Instrumentation and Detectors, 
 ed. M. Iye\& A.F., Moorwood, ESO, Vol. 4008, 1132
\bibitem[Shu(1991)]{Shu} Shu, F. H. 1991, in The Physics of Star Formation and Early
Stellar Evolution, ed. C. J. Lada and N. D. Kylafis (Dordrecht: Kluwer), 365
\bibitem[Tinney(1998)]{Tinney} Tinney, C.C. 1998, \mnras, 296, L42
\end{thebibliography}
\end{document}